\documentclass[aps,prl,reprint,superscriptaddress,nofootinbib,longbibliography]{revtex4-2}

\usepackage[T1]{fontenc}
\usepackage[utf8]{inputenc}
\usepackage{amsmath,amssymb,bm,mathtools}
\usepackage{graphicx}
\usepackage{microtype}
\usepackage{hyperref}
\usepackage{xcolor}
\usepackage{comment}
\hypersetup{colorlinks=true,citecolor=blue,linkcolor=blue,urlcolor=blue}
\usepackage[percent]{overpic}
\newcommand{\vac}{\lvert 0\rangle}
\newcommand{\ket}[1]{\lvert #1\rangle}

\newcommand{\nb}{n_{\mathrm b}}

\begin{document}

\title{Programmable photonic state fusion via heralded storage of asynchronously generated resources}

\author{Mustafa G\"undo\u{g}an}
\email{mustafa.guendogan@physik.hu-berlin.de}
\affiliation{Institut f\"ur Physik and Center for the Science of Materials Berlin (CSMB), Humboldt-Universit\"at zu Berlin, Berlin 12489, Germany}

\author{Dennis R\"atzel}
\email{dennis.ratzel@ucl.ac.uk}
\affiliation{Department of Physics and Astronomy, University College London, Gower Street, WC1E 6BT, London, United Kingdom}
\affiliation{Vienna Center for Quantum Science and Technology, Atominstitut, TU Wien, Stadionallee 2, 1020 Vienna, Austria}

\date{September 9, 2026}

\begin{abstract}
Probabilistic photonic sources generate elementary states in different trials, whereas multiphoton protocols require them to interfere in common temporal modes. We propose a fusion protocol that overcomes this mismatch by successively loading independently heralded photonic states into active storage loops. Conditioning on vacuum in monitored dump modes selects events in which each newly generated state is transferred into the same circulating modes as the photons already stored, thereby removing its generation time label. Provided the two alternatives of each elementary state undergo the same loading transformation, the accumulated state is described by a product of programmable linear factors, allowing a target superposition to be constructed by polynomial factorization. Adjusting the storage loop coupling as the state grows substantially improves the loading efficiency, changing the faster-than-exponential penalty of fixed balanced couplers to exponential scaling. We apply the protocol to two-photon path--frequency states for photonic clock interferometry and estimate the detected rate including source waiting time and round-trip loss.
\end{abstract}
\maketitle

Photonic fusion joins small entangled resources into larger states and is central to optical quantum information~\cite{Browne2005,Kok2007}. Loss-tolerant, ballistic, percolation-based, and fusion-based architectures have been proposed~\cite{Kieling2007,GimenoSegovia2015,Pant2019,Bartolucci2023}, and recent experiments have demonstrated heralded GHZ resources and boosted fusion gates~\cite{Cao2024,Guo2026}. In many platforms, however, the elementary resources are generated probabilistically and become available in different clock cycles. For a pulsed spontaneous parametric down-conversion (SPDC) source, for example, each pump pulse constitutes a source trial and a herald identifies the trials in which the desired resource was produced. Active switching and temporal multiplexing can collect such events from different time bins~\cite{Pittman2002,Kaneda2015,Kaneda2019,MeyerScott2022,Hou2023}, while optical buffers and quantum memories can synchronize them and increase multiphoton generation rates~\cite{Nunn2013,Kaneda2017}. This does not in general place the fused photons in common temporal modes: the surviving state may still retain information about the generation times of its constituent resources. For the states considered here, these labels must be removed so that amplitudes associated with different generation histories can interfere.

We consider a sequential architecture in which heralded fixed-photon-number signal resources, which we call \emph{blocks}, are loaded into active storage loops. The aim is to combine blocks generated in different source trials into a multiphoton state occupying common temporal modes. The source and loop settings assigned to a loading step are held until the required herald is obtained, after which the block is routed to the loop couplers. Monitoring the dump modes selects the branch in which the incoming photons are transferred into the same circulating modes as the photons already stored. In this branch, the original time-bin label is removed and the newly loaded block becomes part of the same stored multiphoton state. A dump event rejects the sequence, whereas a successful load advances the controller to the next setting. After $N$ successful loads, the accumulated state is released in a common output time.

Each block is prepared as a coherent superposition of two logical alternatives. We require the corresponding storage modes to undergo the same complex loading transformation at every step, including their spectral, temporal, and phase response. Under this condition, the amplitude for a successful load depends on the loading step but not on the logical composition of the state already stored. Successive loading operations therefore produce a product of linear factors whose coefficients are controlled by the elementary blocks. This gives a direct prescription for state synthesis: factorization of the polynomial associated with the desired output state determines the block amplitudes and phases. The storage couplings determine the probability of assembling that state, and adapting them as the stored photon number increases changes the loading overhead from faster-than-exponential for fixed balanced couplers to exponential scaling.

Our motivating application is a clock-interferometry state assembled from two-photon blocks~\cite{Barzel2024,GundoganClock2026}. Each block contains one photon in each of two spatial modes, while its two logical alternatives exchange the frequency assignment between them. The elementary resources are therefore themselves composite multiphoton states. We first formulate the fusion process for an arbitrary fixed block photon number $\nb$, without specifying how the blocks are generated. We then consider the two-photon clock-state realization, including an idler Bell-state measurement for block preparation, the required matching of the storage loops, and the effect of source waiting time and storage loss on the output rate.

\emph{General fusion process.--}
The source and loop settings are indexed by successful loading events. Thus $j=0,\ldots,N-1$ labels the loading step rather than the source trial, and the block loaded at step $j$ is
\begin{equation}
\ket{\psi_j}=\left(u_jX_j^\dagger+v_jY_j^\dagger\right)\vac,
\qquad |u_j|^2+|v_j|^2=1.
\label{eq:block}
\end{equation}
For each $j$, $X_j^\dagger\vac$ and $Y_j^\dagger\vac$ denote normalized, orthogonal $\nb$-photon states. The operators $X_j^\dagger$ and $Y_j^\dagger$ may be homogeneous polynomials of creation operators and may therefore describe entangled multiphoton configurations. They commute because they contain only bosonic creation operators. We assume that the loading index changes only the incoming time bin: for fixed $Z=X,Y$, $Z_j^\dagger$ and $Z_k^\dagger$ create the same photon configuration at different source times. Once the block has entered the common storage modes, we denote the corresponding operator by $Z^\dagger$ and omit the time index.

\begin{figure}[t]
\centering

\raisebox{0cm}{%
\begin{overpic}[width=0.62\columnwidth,height=3cm]{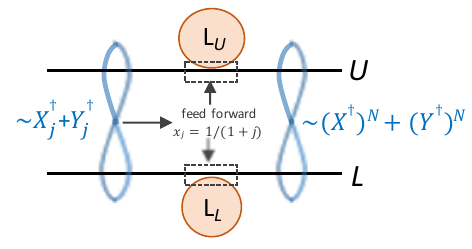}
    \put(2,55){\large\textbf{a)}}
\end{overpic}%
}
\hspace{0.02\columnwidth}
\raisebox{0cm}{%
\begin{overpic}[width=0.33\columnwidth,height=2.7cm]{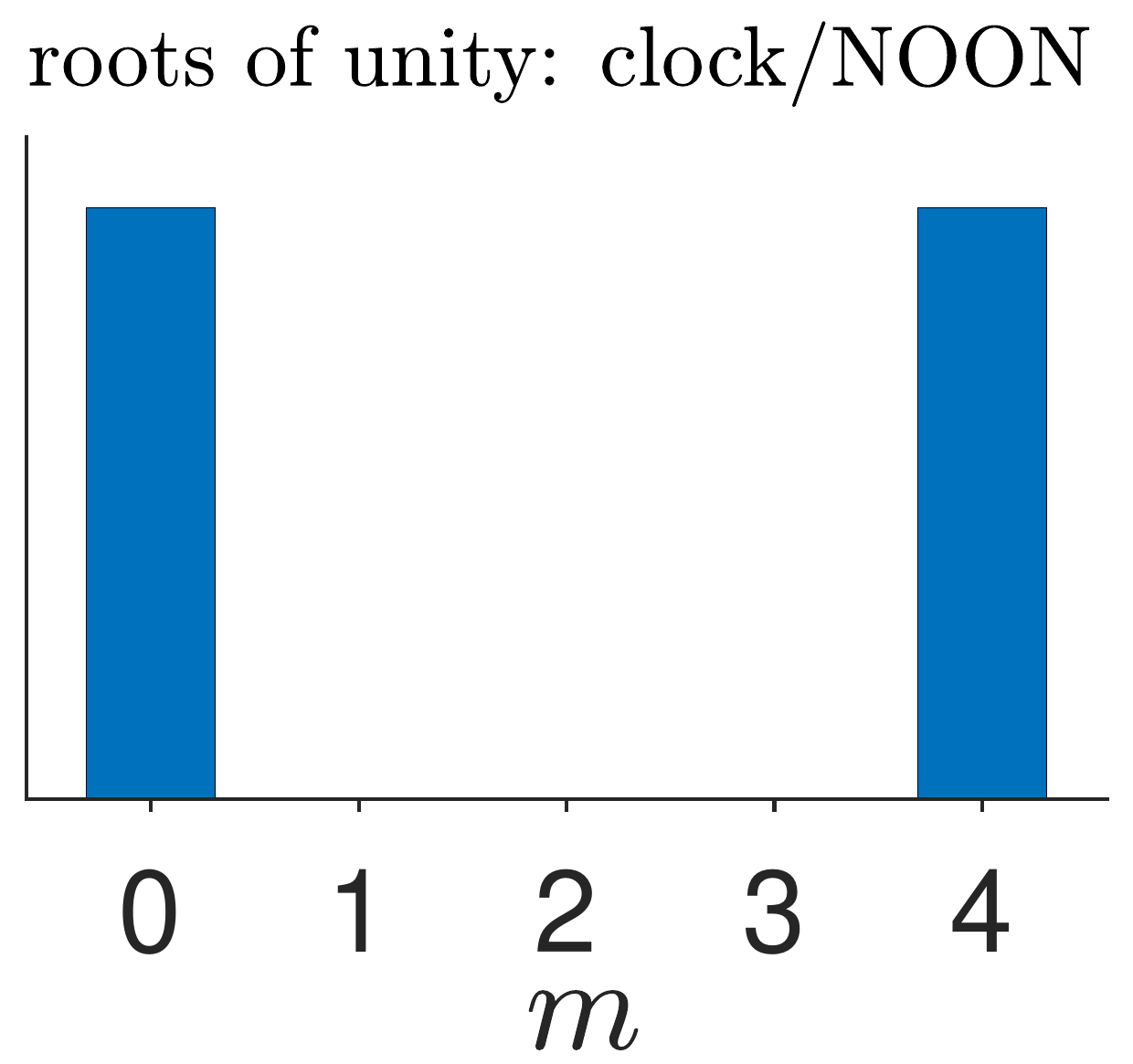}
    \put(-20,95){\large\textbf{b)}}
\end{overpic}%
}

\caption{Bosonic fusion for the clock-state implementation.
    (a) A specific realization of our fusion scheme for generating the photonic clock state  considered for clock interferometry in~\cite{Barzel2024,GundoganClock2026}. Once an elementary block $\sim X_j^\dagger+Y_j^\dagger$ is heralded, the herald is fed forward to the tunable loop couplers, which are set to the programmed coupling $x_j$. Successful loading places the newly generated photons into the same circulating modes as the previously stored photons, removing their generation-time labels. Repeating the procedure for $N$ successful loads yields the fused output.
   (b) Mode-occupation amplitudes obtained with the roots-of-unity phase settings for the case of $N=4$. The index $m$ labels terms proportional to $(X^\dagger)^{N-m}(Y^\dagger)^m$, i.e., configurations with $m$ photons in $Y$ and $N-m$ photons in $X$. Destructive interference cancels all intermediate terms $0<m<N$, leaving only the two components with all $N$ photons in $X$ or all $N$ photons in $Y$, yielding the clock NOON state $\sim (X^\dagger)^N+(Y^\dagger)^N$.}
\label{fig:architecture}
\end{figure}

For the clock application, let $U,L$ label two spatial modes and $1,2$ two frequency bins. The two alternatives are
\begin{equation}
X_j^\dagger=a_{U,1,j}^\dagger a_{L,2,j}^\dagger,
\qquad
Y_j^\dagger=a_{U,2,j}^\dagger a_{L,1,j}^\dagger.
\label{eq:clockXY}
\end{equation}
Each alternative contains one photon in $U$ and one in $L$, with the two frequency assignments exchanged. If both $u_j$ and $v_j$ are nonzero, the block state in Eq.~\eqref{eq:block} is frequency entangled. A successful load maps the incoming operators to $X^\dagger=a_{U,1}^\dagger a_{L,2}^\dagger$ and $Y^\dagger=a_{U,2}^\dagger a_{L,1}^\dagger$ in the common storage modes.
The sequence is shown in Fig.~\ref{fig:architecture}(a).

After $N$ loads, the stored state lies in the span of the $N+1$ monomials containing $m$ copies of $Y^\dagger$ and $N-m$ copies of $X^\dagger$. We write the desired state in this subspace as
\begin{equation}
\begin{aligned}
\ket{\Psi_{\mathrm{tar}}}&=\sum_{m=0}^{N}d_m\ket{m;N}_{XY},\\
\ket{m;N}_{XY}&=\frac{(X^\dagger)^{N-m}(Y^\dagger)^m}{\nu_m}\vac,
\end{aligned}
\label{eq:target}
\end{equation}
where $m$ counts the number of $Y$ blocks and $\nu_m$ is the norm of the corresponding unnormalized monomial.
\footnote{If the normalized monomial states are not mutually orthogonal, their Gram matrix enters the normalization of the superposition.}
We assume that these monomials are linearly independent, so the coefficients $d_m$ are unique. Their span defines the target subspace and need not coincide with the full Hilbert space of the underlying optical modes.

The map describing the coupling into the storage and the post-selection via the monitoring of the dump mode may be stated through a Kraus operator $K_j$. The target state can be reached if, given the states $|\Phi_m^{(j)}\rangle=(X^\dagger)^{j-m}(Y^\dagger)^m|0\rangle$, the Kraus operators $K_j$ satisfy the condition

\begin{equation}
\label{eq:krauscondition}
K_j Z^\dagger_j |\Phi_m^{(j)}\rangle = \alpha_j Z^\dagger|\Phi_m^{(j)}\rangle,
\qquad
Z\in{X,Y},\quad m=0,\ldots,j.
\end{equation}

\noindent with $\alpha_j$ independent of $m$ and $Z$. Under this condition, the same complex amplitude $\alpha_j$ applies to both logical alternatives and to every preceding logical history. Identical spectral, temporal, and phase response of each matched pair of $X$ and $Y$ modes is a sufficient condition. A known differential phase may be included in $v_j/u_j$, whereas a fluctuating phase changes the prepared state. Repeated application of Eq.~\eqref{eq:krauscondition} gives
\begin{equation}
\ket{\widetilde\Psi_N}
=\alpha\prod_{j=0}^{N-1}\left(u_jX^\dagger+v_jY^\dagger\right)\vac,
\qquad \alpha=\prod_{j=0}^{N-1}\alpha_j.
\label{eq:storedproduct}
\end{equation}
All histories with the same number of $X$ and $Y$ blocks now interfere coherently. Defining the polynomial 
\begin{equation}
Q_{\mathrm{tar}}(\theta)
\equiv\sum_{m=0}^{N}\frac{d_m}{\nu_m}\theta^m,
\label{eq:targetpoly}
\end{equation}
the necessary setting to reach the target state follows from the condition
\begin{equation}
Q_{\mathrm{tar}}(\theta)=q\prod_{j=0}^{N-1}(u_j+v_j\theta).
\label{eq:factorization}
\end{equation}
This condition states that every finite nonzero root of $Q_{\mathrm{tar}}(\theta)$ fixes one ratio $v_j/u_j$. Furthermore, a global factor $\theta^l$ in $Q_{\mathrm{tar}}(\theta)$ corresponds to vanishing coefficients $d_m$ for $m<l$ and is implemented by $l$ pure $Y$ blocks. Analogously, if $d_m$ vanishes for all $m>N-l$, $l$ pure $X$ blocks have to be added. Thus amplitude and phase control of the elementary blocks prepares every pure state in the target subspace defined by Eq.~\eqref{eq:target}. Appendix~\ref{app:polynomial} gives the coefficient expansion and its relation to the Majorana representation~\cite{Bastin2009,Devi2012}.

An implementation of the loading sequence may be given when the block operators are monomials of creation operators associated with normalized, mutually orthogonal single-photon modes,
\begin{equation}
X_j^\dagger=\prod_{\ell=1}^{\nb}a_{\ell,X,j}^\dagger,
\qquad
Y_j^\dagger=\prod_{\ell=1}^{\nb}a_{\ell,Y,j}^\dagger.
\label{eq:monomialblocks}
\end{equation}
At step $j$, let $s_{\ell j}$ be the complex amplitude for the incoming photon labelled by $\ell$ to enter its storage mode, and let $c_{\ell j}$ be the amplitude for a photon already occupying that mode to remain stored. We use the same $s_{\ell j}$ and $c_{\ell j}$ for the matched $X$ and $Y$ modes. The condition in Eq.~\eqref{eq:krauscondition} is then satisfied with
\begin{equation}
\alpha_j=\prod_{\ell=1}^{\nb}s_{\ell j}c_{\ell j}^{\,j}.
\label{eq:alphaj}
\end{equation}
The power $c_{\ell j}^{j}$ depends only on the total number of stored blocks, not on their distribution between $X$ and $Y$. Hence the same $\alpha_j$ applies to every logical history.

A particularly interesting state can be reached by setting $u_j=1/\sqrt2$ and $v_j=z_0\omega^j/\sqrt2$, where $\omega=e^{-2\pi i/N}$ and $|z_0|=1$. The identity
\begin{equation}
\prod_{j=0}^{N-1}\left(X^\dagger+z_0\omega^jY^\dagger\right)
=(X^\dagger)^N+(-1)^{N+1}z_0^N(Y^\dagger)^N
\label{eq:rootsunity}
\end{equation}
shows that every term containing both alternatives cancels. We refer to these phase settings as the roots of unity setting. The two surviving terms, $(X^\dagger)^N$ and $(Y^\dagger)^N$, corresponding to $m=0$ and $m=N$, respectively, lie at the two ends of the occupation expansion and will therefore be referred to as the two endpoints. Their coherent superposition forms a NOON-type state. Choosing $(-1)^{N+1}z_0^N=e^{i\Phi_N}$, we find the stored state
\begin{equation}
\ket{\Psi_N}=\frac{(X^\dagger)^N+e^{i\Phi_N}(Y^\dagger)^N}
{\sqrt{2}(N!)^{\nb/2}}\vac,
\label{eq:endpointstate}
\end{equation}
up to a global phase. 
For the specific choice of the block operators defined in Eq.~\eqref{eq:clockXY}, Eq.~\eqref{eq:endpointstate} gives the bunched path and frequency entangled state used in memory assisted photonic clock interferometry. This protocol was analyzed in Ref.~\cite{Barzel2024}, and its multiphoton extension in Ref.~\cite{GundoganClock2026}. 

For an ideal lossless coupler, $|c_{\ell j}|^{2}=(1-|s_{\ell j}|^2)$, and the probability to retain all $Nn_b$ photons in the storage modes is proportional to
\begin{equation}
|\alpha|^2=\prod_{j=0}^{N-1}\prod_{\ell=1}^{\nb}
x_{\ell,j}(1-x_{\ell,j})^{j}.
\label{eq:alphaall}
\end{equation}
where $x_{\ell,j}=|s_{\ell j}|^2$. Beyond this condition, the amplitudes can be chosen freely. This implies that we can find a maximal $|\alpha|^2$ by maximizing all factors $x_{\ell,j}(1-x_{\ell,j})^{j}$ individually. We conclude that the optimal amplitudes fulfill $x_{\ell,j}=x_j=1/(1+j)$. For the roots of unity setting, we find the loading probability 

\begin{equation}
\begin{aligned}
p_{N,\nb}^{\mathrm{opt,ROU}}
&=2^{1-N}\left[\prod_{j=1}^{N-1}\left(\frac{j}{1+j}\right)^j\right]^{\nb}\\
&=2^{1-N}\left(\frac{(N-1)!}{N^{N-1}}\right)^{\nb}.
\end{aligned}
\label{eq:optimal}
\end{equation}
Using Stirling's formula, $p_{N,\nb}^{\mathrm{opt,ROU}}$ decreases exponentially with $N$, apart from a polynomial prefactor. A fixed balanced coupler, $x_{\ell,j}=1/2$, instead gives a faster-than-exponential decrease. Appendices~\ref{app:roots} and~\ref{app:optimal} give the endpoint normalization and the derivation of Eq.~\eqref{eq:optimal}.

\emph{Two-photon clock-state application.--}
One possible source for the blocks defined by Eq.~\eqref{eq:clockXY} uses two entangled-pair SPDC sources, which provide signal--idler states~\cite{Kwiat1995,Couteau2018}. A selected Bell projection on the idlers realizes entanglement swapping~\cite{Zukowski1993,Pan1998}, and a partial linear-optical Bell analyzer can herald the required outcome~\cite{Calsamiglia2001}. A selected Bell-state measurement (BSM) on the idlers realizes entanglement swapping~\cite{Zukowski1993,Pan1998}, and a partial linear-optical Bell analyzer can herald the required outcome~\cite{Calsamiglia2001}. The corresponding signal photons are left in a coherent superposition of the two exchanged frequency assignments and are routed to $U$ and $L$. For each loading index $j$, the source amplitudes, relative phase, and loop coupling are kept fixed over repeated pump cycles and are changed only after the block has been loaded successfully. Appendix~\ref{app:block-preparation} gives an effective source description.

The storage response must be matched pairwise. Modes $(U,1)$ and $(U,2)$ must experience the same complex transformation at each loading step, and the same requirement applies to $(L,1)$ and $(L,2)$. The common transformation in $U$ may differ from that in $L$. Under these conditions the loops do not acquire information about the exchanged frequency assignment, and Eq.~\eqref{eq:krauscondition} holds for the blocks defined by Eq.~\eqref{eq:clockXY}.

The corresponding NOON-type state in Eq.~\eqref{eq:endpointstate} is the bunched path- and frequency-entangled state used in memory-assisted photonic clock interferometry. The two-photon protocol was analyzed in Ref.~\cite{Barzel2024}, and the multiphoton extension in Ref.~\cite{GundoganClock2026}. In an ordinary path NOON state, the two components correspond to all photons occupying one spatial mode or the other. In contrast, each endpoint of the state generated here contains $N$ photons in each spatial mode, with the two endpoints distinguished by the exchanged assignment of the two frequency bins to the paths.

\emph{Performance.--}
After $k$ successful loads, $k\nb$ photons remain in the loops while the source is operated until the next usable block is heralded. The detected rate is therefore set by the competition between block supply and round-trip loss. For a sequential pulsed source, let $f_{\mathrm{rep}}$ denote the pulse repetition rate, $p_h$ the probability per cycle that a usable $\nb$-photon block reaches the loading stage, $\lambda$ the loss per stored photon in one loop round trip, and $\eta$ the transmission and detection efficiency of each delivered photon. For $p_h,\lambda\ll1$, the mean survival probability during assembly is 
\begin{equation}
S_{N,\nb}\simeq\prod_{k=1}^{N-1}\frac{p_h}{p_h+k\nb\lambda},
\label{eq:survival}
\end{equation}
and an estimate for a lower bound on the detected event rate within this sequential model is
\begin{equation}
R_{\mathrm{out},N}\simeq\frac{f_{\mathrm{rep}}p_h}{N}
S_{N,\nb}\,p_{N,\nb}^{\mathrm{opt,ROU}}\,\eta^{N\nb},
\label{eq:rate}
\end{equation}
where the first factor is the rate of $N$ usable blocks delivered to the storage coupler per second. $R_{\mathrm{out},N}$ is only a lower bound because one may restart the sequence after an early failed attempt which would be detected by observing photons in the dump port. The waiting time average leading to Eqs.~\eqref{eq:survival} and \eqref{eq:rate} is derived in Appendix~\ref{app:waiting}.

\begin{figure}[t]
\centering
\begin{overpic}[width=0.90\columnwidth]{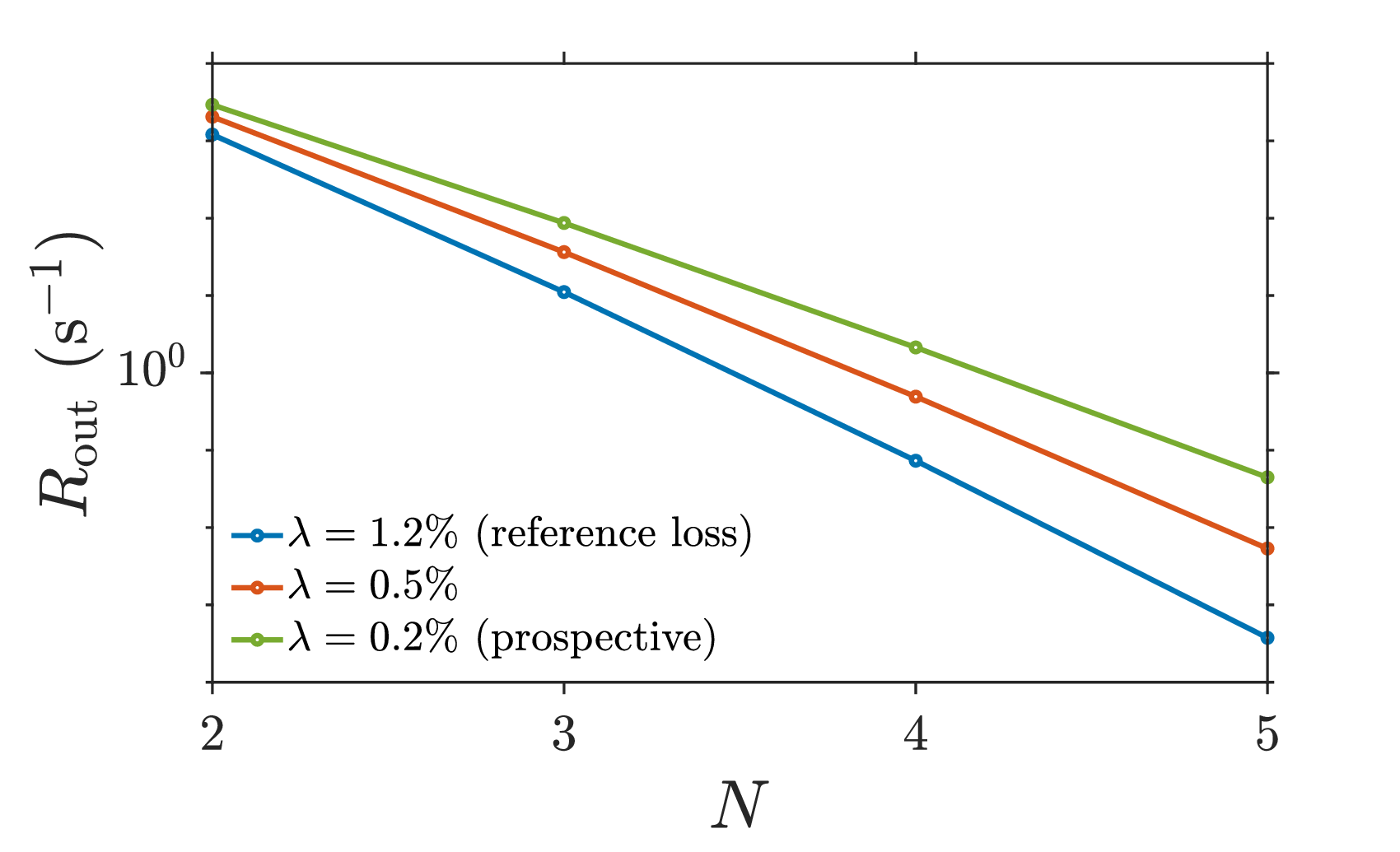}
    \put(0,57){\large\textbf{a)}}
\end{overpic}

\begin{overpic}[width=0.90\columnwidth]{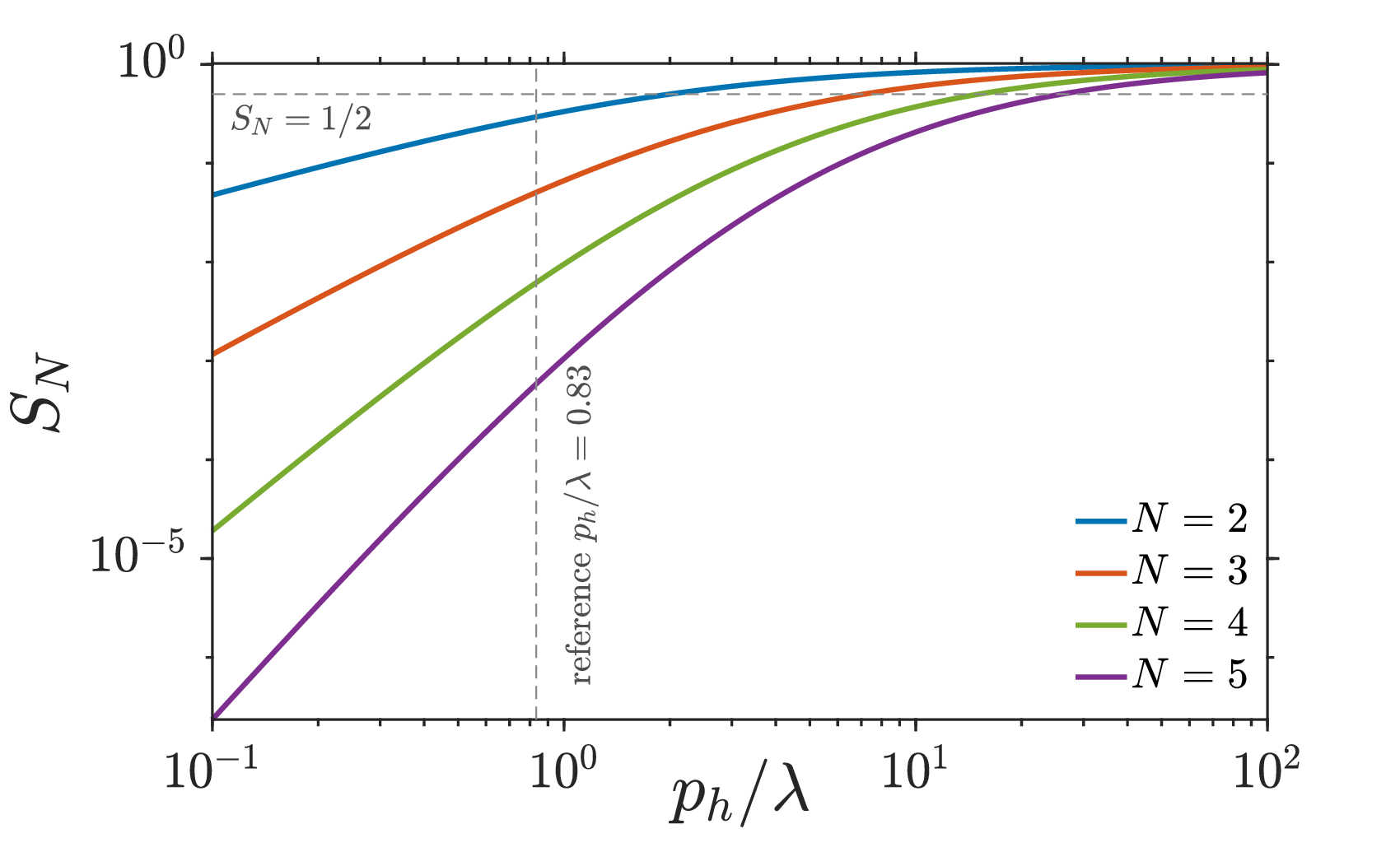}
    \put(0,57){\large\textbf{b)}}
\end{overpic}

\caption{Performance of the $\nb=2$ path--frequency realization. (a) Detected rates from Eq.~\eqref{eq:rate} for $f_{\rm rep}=10\,\mathrm{MHz}$, $p_h=10^{-2}$, $\eta=0.9$, and ideal dump-port monitoring. The curves correspond to round-trip losses of $1.2\%$, $0.5\%$, and $0.2\%$ per stored photon. The $1.2\%$ value is a reference loss rather than a demonstrated end-to-end operating point. (b) Waiting survival $S_{N,2}$ as a function of $p_h/\lambda$ for $N=2,\ldots,5$. The dashed lines mark $S_{N,2}=1/2$ and the reference ratio $p_h/\lambda\simeq0.83$.}
\label{fig:performance}
\end{figure}

For the reference round-trip loss $\lambda=1.2\%$, Eq.~\eqref{eq:rate} gives $1.2\times10^3\,\mathrm{s}^{-1}$ for $N=2$ and $11\,\mathrm{s}^{-1}$ for $N=3$, as shown in Fig.~\ref{fig:performance}(a). With the same source and detection parameters, a rate of $1\,\mathrm{s}^{-1}$ at $N=4$ requires $\lambda\simeq3.4\times10^{-3}$. The survival calculation in Fig.~\ref{fig:performance}(b) places the reference ratio $p_h/\lambda\simeq0.83$ in the storage-loss-limited regime. 

Reconfigurable loops and photonic buffers provide the repeated loading operations required here~\cite{Motes2014,Evans2023,Takeda2019,Okuno2024}, and iterated photon addition in a loop has been studied in Ref.~\cite{Mendei2024}. Furthermore, frequency-bin entanglement and coherent processing are established experimentally in integrated and microresonator platforms~\cite{Kues2017,Lu2023}. Ordinary two-mode path NOON states can be prepared by other methods, including tailored Fock inputs and heralded passive interferometers~\cite{Armezzani2026}. These works establish the main storage, switching, and frequency-processing ingredients required by the present scheme. The present protocol allows each incoming resource to contain several photons and uses the matched complex response to keep the transfer coefficient independent of its logical composition. Related schemes for symmetric-state preparation use conditioned photon addition, stationary qubits, global control, or general multimode photonics~\cite{McCusker2009,Lamata2013,Bond2025,Aralov2026}.
Projective fusion can also prepare related states, but for the frequency-bin encoding considered here it requires coherent rotations between bins 1 and 2 before Bell or parity detection~\cite{Lu2018,Lingaraju2022}. Such frequency-domain Bloch-basis measurements generally require active frequency conversion or equivalent electro-optic mixing of the individual frequency components. Our architecture avoids this requirement altogether: the Bell projection is performed only on the idlers that herald the elementary blocks, while the delivered signal photons undergo neither Bell measurements nor frequency-bin rotations during assembly. Consequently, the encoded signal modes can remain spectrally untouched throughout the fusion sequence, with the implementation relying only on mode-matched loop coupling, phase stability, and low-loss storage.
The same architecture naturally extends to larger fusion sequences and to multiplexed block sources, where parallel generation can suppress stochastic waiting times and reduce the required storage depth. More generally, the matched-map condition is not tied to a particular loop realization and can be implemented with other optical buffers or quantum-memory interfaces. This opens a route to assembling asynchronously generated entangled resources across different photonic encodings while preserving the delivered photons for subsequent sensing, communication, or coherent processing.

\emph{Acknowledgments,--}
MG acknowledges support from the Einstein Foundation Berlin through an Independent Researcher Grant and  support from DLR through funds provided by BMFTR (OPTIMUS PRIME, No. 50SI2655A). DR acknowledges financial support from EPSRC (Engineering \& Physical Sciences Research Council, United Kingdom) Grant Number EP/X009467/1, and support by the Deutsche Forschungsgemeinschaft (DFG, German Research Foundation) under Germany’s Excellence Strategy – EXC-2123 QuantumFrontiers – 390837967.

\bibliography{references}

@article{Browne2005,
  author  = {Browne, Daniel E. and Rudolph, Terry},
  title   = {Resource-efficient linear optical quantum computation},
  journal = {Phys. Rev. Lett.},
  volume  = {95},
  pages   = {010501},
  year    = {2005},
  doi     = {10.1103/PhysRevLett.95.010501}
}

@article{Kok2007,
  author  = {Kok, Pieter and Munro, William J. and Nemoto, Kae and Ralph, Timothy C. and Dowling, Jonathan P. and Milburn, Gerard J.},
  title   = {Linear optical quantum computing with photonic qubits},
  journal = {Rev. Mod. Phys.},
  volume  = {79},
  pages   = {135--174},
  year    = {2007},
  doi     = {10.1103/RevModPhys.79.135}
}

@article{Kieling2007,
  author  = {Kieling, K. and Rudolph, T. and Eisert, J.},
  title   = {Percolation, renormalization, and quantum computing with nondeterministic gates},
  journal = {Phys. Rev. Lett.},
  volume  = {99},
  pages   = {130501},
  year    = {2007},
  doi     = {10.1103/PhysRevLett.99.130501}
}

@article{GimenoSegovia2015,
  author  = {Gimeno-Segovia, Mercedes and Shadbolt, Peter and Browne, Dan E. and Rudolph, Terry},
  title   = {From three-photon {Greenberger--Horne--Zeilinger} states to ballistic universal quantum computation},
  journal = {Phys. Rev. Lett.},
  volume  = {115},
  pages   = {020502},
  year    = {2015},
  doi     = {10.1103/PhysRevLett.115.020502}
}

@article{Pant2019,
  author  = {Pant, Mihir and Towsley, Don and Englund, Dirk and Guha, Saikat},
  title   = {Percolation thresholds for photonic quantum computing},
  journal = {Nat. Commun.},
  volume  = {10},
  pages   = {1070},
  year    = {2019},
  doi     = {10.1038/s41467-019-08948-5}
}

@article{Bartolucci2023,
  author  = {Bartolucci, Sara and Birchall, Patrick and Bombin, Hector and Cable, Hugo and Dawson, Chris and Gimeno-Segovia, Mercedes and Johnston, Eric and Kieling, Kamil and Nickerson, Naomi and Pant, Mihir and Pastawski, Fernando and Rudolph, Terry and Sparrow, Chris},
  title   = {Fusion-based quantum computation},
  journal = {Nat. Commun.},
  volume  = {14},
  pages   = {912},
  year    = {2023},
  doi     = {10.1038/s41467-023-36493-1}
}

@article{Cao2024,
  author  = {Cao, H. and Hansen, L. M. and Giorgino, F. and Carosini, L. and Zahalka, P. and Zilk, F. and Loredo, J. C. and Walther, P.},
  title   = {Photonic source of heralded {Greenberger--Horne--Zeilinger} states},
  journal = {Phys. Rev. Lett.},
  volume  = {132},
  pages   = {130604},
  year    = {2024},
  doi     = {10.1103/PhysRevLett.132.130604}
}

@article{Guo2026,
  author  = {Guo, Yong-Peng and Zou, Geng-Yan and Ding, Xing and Zhang, Qi-Hang and Xu, Mo-Chi and Liu, Run-Ze and Zhao, Jun-Yi and Ge, Zhen-Xuan and Peng, Li-Chao and Xu, Ke-Mi and Lou, Yi-Yang and Ning, Zhen and Wang, Lin-Jun and Wang, Hui and Huo, Yong-Heng and He, Yu-Ming and Lu, Chao-Yang and Pan, Jian-Wei},
  title   = {Boosted fusion gates above the percolation threshold for scalable graph-state generation},
  journal = {Phys. Rev. A},
  volume  = {113},
  pages   = {L040602},
  year    = {2026},
  doi     = {10.1103/PhysRevA.113.L040602}
}

@article{Pittman2002,
  author  = {Pittman, T. B. and Franson, J. D.},
  title   = {Cyclical quantum memory for photonic qubits},
  journal = {Phys. Rev. A},
  volume  = {66},
  pages   = {062302},
  year    = {2002},
  doi     = {10.1103/PhysRevA.66.062302}
}

@article{Kaneda2015,
  author  = {Kaneda, Fumihiro and Christensen, Bradley G. and Wong, Jia Jun and Park, Heonoh S. and McCusker, Kevin T. and Kwiat, Paul G.},
  title   = {Time-multiplexed heralded single-photon source},
  journal = {Optica},
  volume  = {2},
  pages   = {1010--1013},
  year    = {2015},
  doi     = {10.1364/OPTICA.2.001010}
}

@article{Kaneda2019,
  author  = {Kaneda, Fumihiro and Kwiat, Paul G.},
  title   = {High-efficiency single-photon generation via large-scale active time multiplexing},
  journal = {Sci. Adv.},
  volume  = {5},
  pages   = {eaaw8586},
  year    = {2019},
  doi     = {10.1126/sciadv.aaw8586}
}

@article{MeyerScott2022,
  author  = {Meyer-Scott, Evan and Prasannan, Nidhin and Dhand, Ish and Eigner, Christof and Quiring, Viktor and Barkhofen, Sonja and Brecht, Benjamin and Plenio, Martin B. and Silberhorn, Christine},
  title   = {Scalable generation of multiphoton entangled states by active feed-forward and multiplexing},
  journal = {Phys. Rev. Lett.},
  volume  = {129},
  pages   = {150501},
  year    = {2022},
  doi     = {10.1103/PhysRevLett.129.150501}
}

@article{Hou2023,
  author  = {Hou, Z. and others},
  title   = {Entangled-state time multiplexing for multiphoton entanglement generation},
  journal = {Phys. Rev. Applied},
  volume  = {19},
  pages   = {L011002},
  year    = {2023},
  doi     = {10.1103/PhysRevApplied.19.L011002}
}

@article{Nunn2013,
  author  = {Nunn, J. and Langford, N. K. and Kolthammer, W. S. and Champion, T. F. M. and Sprague, M. R. and Michelberger, P. S. and Jin, X.-M. and England, D. G. and Walmsley, I. A.},
  title   = {Enhancing multiphoton rates with quantum memories},
  journal = {Phys. Rev. Lett.},
  volume  = {110},
  pages   = {133601},
  year    = {2013},
  doi     = {10.1103/PhysRevLett.110.133601}
}

@article{Kaneda2017,
  author  = {Kaneda, Fumihiro and Xu, Feihu and Chapman, Joseph and Kwiat, Paul G.},
  title   = {Quantum-memory-assisted multiphoton generation for efficient quantum information processing},
  journal = {Optica},
  volume  = {4},
  pages   = {1034--1037},
  year    = {2017},
  doi     = {10.1364/OPTICA.4.001034}
}

@article{Barzel2024,
  author  = {Barzel, Roy and G{\"u}ndo{\u{g}}an, Mustafa and Krutzik, Markus and R{\"a}tzel, Dennis and L{\"a}mmerzahl, Claus},
  title   = {Entanglement dynamics of photon pairs and quantum memories in the gravitational field of the {Earth}},
  journal = {Quantum},
  volume  = {8},
  pages   = {1273},
  year    = {2024},
  doi     = {10.22331/q-2024-02-29-1273}
}

@misc{GundoganClock2026,
  author        = {G{\"u}ndo{\u{g}}an, Mustafa and Barzel, Roy and R{\"a}tzel, Dennis},
  title         = {Gravitational time dilation in quantum clock interferometry with entangled multi-photon states and quantum memories},
  year          = {2026},
  eprint        = {2601.02470},
  archivePrefix = {arXiv},
  primaryClass  = {quant-ph}
}

@article{Kwiat1995,
  author  = {Kwiat, Paul G. and Mattle, Klaus and Weinfurter, Harald and Zeilinger, Anton and Sergienko, Alexander V. and Shih, Yanhua},
  title   = {New high-intensity source of polarization-entangled photon pairs},
  journal = {Phys. Rev. Lett.},
  volume  = {75},
  pages   = {4337--4341},
  year    = {1995},
  doi     = {10.1103/PhysRevLett.75.4337}
}

@article{Couteau2018,
  author  = {Couteau, Christophe},
  title   = {Spontaneous parametric down-conversion},
  journal = {Contemp. Phys.},
  volume  = {59},
  pages   = {291--304},
  year    = {2018},
  doi     = {10.1080/00107514.2018.1488463}
}

@article{Zukowski1993,
  author  = {{\.{Z}}ukowski, Marek and Zeilinger, Anton and Horne, Michael A. and Ekert, Artur K.},
  title   = {``Event-ready-detectors'' {Bell} experiment via entanglement swapping},
  journal = {Phys. Rev. Lett.},
  volume  = {71},
  pages   = {4287--4290},
  year    = {1993},
  doi     = {10.1103/PhysRevLett.71.4287}
}

@article{Pan1998,
  author  = {Pan, Jian-Wei and Bouwmeester, Dik and Weinfurter, Harald and Zeilinger, Anton},
  title   = {Experimental entanglement swapping: Entangling photons that never interacted},
  journal = {Phys. Rev. Lett.},
  volume  = {80},
  pages   = {3891--3894},
  year    = {1998},
  doi     = {10.1103/PhysRevLett.80.3891}
}

@article{Calsamiglia2001,
  author  = {Calsamiglia, John and L{\"u}tkenhaus, Norbert},
  title   = {Maximum efficiency of a linear-optical {Bell}-state analyzer},
  journal = {Appl. Phys. B},
  volume  = {72},
  pages   = {67--71},
  year    = {2001},
  doi     = {10.1007/s003400000484}
}

@article{Kues2017,
  author  = {Kues, Michael and Reimer, Christian and Roztocki, Piotr and Cortes, Luis Romero and Sciara, Stefania and Wetzel, Benjamin and Zhang, Yanbing and Cino, Alfonso and Chu, Sai T. and Little, Brent E. and Moss, David J. and Caspani, Lucia and Aza{\~n}a, Jos{\'e} and Morandotti, Roberto},
  title   = {On-chip generation of high-dimensional entangled quantum states and their coherent control},
  journal = {Nature},
  volume  = {546},
  pages   = {622--626},
  year    = {2017},
  doi     = {10.1038/nature22986}
}

@article{Lu2023,
  author  = {Lu, Hsuan-Hao and Liscidini, Marco and Gaeta, Alexander L. and Weiner, Andrew M. and Lukens, Joseph M.},
  title   = {Frequency-bin photonic quantum information},
  journal = {Optica},
  volume  = {10},
  pages   = {1655--1671},
  year    = {2023},
  doi     = {10.1364/OPTICA.497171}
}

@misc{Armezzani2026,
  author        = {Armezzani, Marcello and Lualdi, Colin P. and Gu, Xuemei and Kwiat, Paul G. and Krenn, Mario},
  title         = {Automated discovery of high-probability heralded schemes for path-entangled states},
  year          = {2026},
  eprint        = {2607.25501},
  archivePrefix = {arXiv},
  primaryClass  = {quant-ph}
}

@article{Motes2014,
  author  = {Motes, Keith R. and Gilchrist, Alexei and Dowling, Jonathan P. and Rohde, Peter P.},
  title   = {Scalable boson sampling with time-bin encoding using a loop-based architecture},
  journal = {Phys. Rev. Lett.},
  volume  = {113},
  pages   = {120501},
  year    = {2014},
  doi     = {10.1103/PhysRevLett.113.120501}
}

@article{Evans2023,
  author  = {Evans, C. J. and Nunn, C. M. and Cheng, S. W. L. and Franson, J. D. and Pittman, T. B.},
  title   = {Experimental storage of photonic polarization entanglement in a broadband loop-based quantum memory},
  journal = {Phys. Rev. A},
  volume  = {108},
  pages   = {L050601},
  year    = {2023},
  doi     = {10.1103/PhysRevA.108.L050601}
}

@article{Takeda2019,
  author  = {Takeda, Shuntaro and Takase, Kan and Furusawa, Akira},
  title   = {On-demand photonic entanglement synthesizer},
  journal = {Sci. Adv.},
  volume  = {5},
  pages   = {eaaw4530},
  year    = {2019},
  doi     = {10.1126/sciadv.aaw4530},
  eprint  = {1811.10704},
  archivePrefix = {arXiv},
  primaryClass = {quant-ph}
}

@article{Okuno2024,
  author  = {Okuno, Daisuke and Yoshida, Takahiro and Arita, Ryuji and Kashiwazaki, Takashi and Umeki, Takeshi and Miki, Shigehito and Terai, Hirotaka and Yabuno, Masahiro and China, Fumihiro and Takeda, Shuntaro},
  title   = {Time-domain programmable beam-splitter operations for an optical phase-sensitive non-{Gaussian} state},
  journal = {Phys. Rev. A},
  volume  = {110},
  pages   = {023706},
  year    = {2024},
  doi     = {10.1103/PhysRevA.110.023706},
  eprint  = {2402.14372},
  archivePrefix = {arXiv},
  primaryClass = {quant-ph}
}

@article{Mendei2024,
  author  = {Mendei, Bal{\'a}zs and Koniorczyk, Matyas and Homa, Gergely and {\'A}d{\'a}m, P{\'e}ter},
  title   = {Photon number states via iterated photon addition in a loop},
  journal = {Photonics},
  volume  = {11},
  pages   = {1075},
  year    = {2024},
  doi     = {10.3390/photonics11111075},
  eprint  = {2406.19207},
  archivePrefix = {arXiv},
  primaryClass = {quant-ph}
}

@article{Bastin2009,
  author  = {Bastin, Thierry and Krins, Stefaan and Mathonet, Pierre and Godefroid, Lucien and Lamata, Lucas and Solano, Enrique},
  title   = {Operational families of entanglement classes for symmetric $N$-qubit states},
  journal = {Phys. Rev. Lett.},
  volume  = {103},
  pages   = {070503},
  year    = {2009},
  doi     = {10.1103/PhysRevLett.103.070503}
}

@article{Devi2012,
  author  = {Devi, A. R. Usha and Sudha and Rajagopal, A. K.},
  title   = {Majorana representation of symmetric multiqubit states},
  journal = {Quantum Inf. Process.},
  volume  = {11},
  pages   = {685--710},
  year    = {2012},
  doi     = {10.1007/s11128-011-0280-8},
  eprint  = {1103.3640},
  archivePrefix = {arXiv},
  primaryClass = {quant-ph}
}

@article{McCusker2009,
  author  = {McCusker, Kevin T. and Kwiat, Paul G.},
  title   = {Efficient optical quantum state engineering},
  journal = {Phys. Rev. Lett.},
  volume  = {103},
  pages   = {163602},
  year    = {2009},
  doi     = {10.1103/PhysRevLett.103.163602},
  eprint  = {0907.1902},
  archivePrefix = {arXiv},
  primaryClass = {quant-ph}
}

@article{Lamata2013,
  author  = {Lamata, Lucas and L{\'o}pez, Carlos E. and Lanyon, Benjamin P. and Bastin, Thierry and Retamal, Juan Carlos and Solano, Enrique},
  title   = {Deterministic generation of arbitrary symmetric states and entanglement classes},
  journal = {Phys. Rev. A},
  volume  = {87},
  pages   = {032325},
  year    = {2013},
  doi     = {10.1103/PhysRevA.87.032325},
  eprint  = {1211.0404},
  archivePrefix = {arXiv},
  primaryClass = {quant-ph}
}

@article{Bond2025,
  author  = {Bond, Liam J. and Davis, M. J. and Min{\'a}{\v{r}}, Ji{\v{r}}{\'i} and Gerritsma, R. and Brennen, G. K. and Safavi-Naini, A.},
  title   = {Global variational quantum circuits for arbitrary symmetric state preparation},
  journal = {Phys. Rev. Research},
  volume  = {7},
  pages   = {L022072},
  year    = {2025},
  doi     = {10.1103/PhysRevResearch.7.L022072}
}

@article{Aralov2026,
  author  = {Aralov, Andrei and Gillet, {\'E}milie and Nguyen, Viet and Cosentino, Andrea and Walschaers, Mattia and Frigerio, Massimo},
  title   = {Photon catalysis for general multimode multi-photon quantum state preparation},
  journal = {PRX Quantum},
  volume  = {7},
  pages   = {020323},
  year    = {2026},
  doi     = {10.1103/ktc9-9rjb}
}

@article{Lu2018,
  author  = {Lu, Hsuan-Hao and Lukens, Joseph M. and Peters, Nicholas A. and Odele, Ogaga D. and Leaird, Daniel E. and Weiner, Andrew M. and Lougovski, Pavel},
  title   = {Electro-optic frequency beam splitters and tritters for high-fidelity photonic quantum information processing},
  journal = {Phys. Rev. Lett.},
  volume  = {120},
  pages   = {030502},
  year    = {2018},
  doi     = {10.1103/PhysRevLett.120.030502}
}

@article{Lingaraju2022,
  author  = {Lingaraju, N. B. and Lu, Hsuan-Hao and Leaird, Daniel E. and Estrella, S. and Lukens, Joseph M. and Weiner, Andrew M.},
  title   = {{Bell} state analyzer for spectrally distinct photons},
  journal = {Optica},
  volume  = {9},
  pages   = {280--283},
  year    = {2022},
  doi     = {10.1364/OPTICA.443302}
}

\appendix
\setcounter{secnumdepth}{1}

\section{Polynomial programming and the symmetric sector}
\label{app:polynomial}

Expanding the product in Eq.~\eqref{eq:storedproduct} gives
\begin{equation}
\prod_{j=0}^{N-1}(u_jX^\dagger+v_jY^\dagger)
=\sum_{m=0}^{N}E_m(X^\dagger)^{N-m}(Y^\dagger)^m,
\label{S-eq:expansion}
\end{equation}
where
\begin{equation}
E_m=\sum_{\substack{S\subseteq\{0,\ldots,N-1\}\\|S|=m}}
\prod_{j\notin S}u_j\prod_{j\in S}v_j.
\label{S-eq:elementary}
\end{equation}
Equivalently, $E_m$ is the coefficient of $\theta^m$ in
\begin{equation}
\prod_{j=0}^{N-1}(u_j+v_j\theta).
\end{equation}
A desired coefficient vector $\{c_m\}$ is therefore obtained by factorizing
\begin{equation}
Q_{\rm tar}(\zeta)=\sum_{m=0}^{N}c_m\theta^m
=q\prod_{j=0}^{N-1}(u_j+v_j\theta).
\label{S-eq:factor}
\end{equation}
Every complex polynomial of degree at most $N$ can be written in this form when pure $X$ and pure $Y$ factors are allowed. A factor $\theta$ sets $u_j=0$, while a factor independent of $\theta$ sets $v_j=0$. After factorization, each pair $(u_j,v_j)$ is normalized and the removed magnitude is absorbed into $q$.

Equation~\eqref{S-eq:expansion} is written in terms of the unnormalized monomials $(X^\dagger)^{N-m}(Y^\dagger)^m\vac$. Define
\begin{equation}
\nu_m=\left\|(X^\dagger)^{N-m}(Y^\dagger)^m\vac\right\|
\end{equation}
and write the target as $\sum_m d_m\nu_m^{-1}(X^\dagger)^{N-m}(Y^\dagger)^m\vac$. The coefficients in Eq.~\eqref{S-eq:factor} are then $c_m=d_m/\nu_m$.  The normalized states $\nu_m^{-1}(X^\dagger)^{N-m}(Y^\dagger)^m\vac$ need not be mutually orthogonal; linear independence is sufficient for the coefficients $c_m$ to be unique. 

For $X^\dagger=a^\dagger$ and $Y^\dagger=b^\dagger$, the resulting $N$-photon state belongs to the symmetric $N$-qubit, or spin-$N/2$, sector. Each factor $u_jX^\dagger+v_jY^\dagger$ specifies one Majorana spinor, and Eq.~\eqref{S-eq:factor} is the occupation-number form of the Majorana constellation~\cite{Bastin2009,Devi2012}. For composite blocks, the polynomial in Eq.~\eqref{S-eq:factor} acts on two logical alternatives that may each contain several photons.

A target containing only the term with $m=M$ is obtained from $M$ pure $Y$ blocks and $N-M$ pure $X$ blocks. In the single-photon realization, this corresponds to a Dicke state and, for $M=N/2$, to a twin-Fock state. 

\section{Roots of unity and endpoint normalization}
\label{app:roots}

Set $u_j=1/\sqrt2$ and $v_j=z_0\omega^j/\sqrt2$, with $|z_0|=1$ and $\omega=e^{2\pi i/N}$. The numbers $-z_0\omega^jY^\dagger$ are the roots of $z^N-(-z_0Y^\dagger)^N$, and hence
\begin{equation}
\prod_{j=0}^{N-1}(X^\dagger+z_0\omega^jY^\dagger)
=(X^\dagger)^N+(-1)^{N+1}z_0^N(Y^\dagger)^N.
\label{S-eq:rou}
\end{equation}
All mixed sectors are absent. For $N=2$, the factors are $X^\dagger+z_0Y^\dagger$ and $X^\dagger-z_0Y^\dagger$. Their mixed terms arise from two source-time histories with opposite signs. Once the second block has been loaded, both histories correspond to the same storage monomial $X^\dagger Y^\dagger$ and cancel.

For the monomial blocks in Eq.~\eqref{eq:monomialblocks}, assume that the $2\nb$ single-photon modes defining $X$ and $Y$ are mutually orthogonal. Then
\begin{equation}
\left\|(X^\dagger)^N\vac\right\|^2
=\left\|(Y^\dagger)^N\vac\right\|^2=(N!)^{\nb}.
\end{equation}
Choosing $(-1)^{N+1}z_0^N=e^{i\Phi_N}$ gives the normalized endpoint state
\begin{equation}
\ket{\Psi_N^{\rm end}}
=\frac{(X^\dagger)^N+e^{i\Phi_N}(Y^\dagger)^N}
{\sqrt2\,(N!)^{\nb/2}}\vac.
\label{S-eq:endstate}
\end{equation}
For the clock realization, $\nb=2$, and Eq.~\eqref{S-eq:endstate} is the state used in the main text.

\section{Optimal coupling schedule and scaling}
\label{app:optimal}

Here $j=0,\ldots,N-1$ labels the successful loading steps, $\nb$ is the number of photons in each elementary block, and $x_j$ is the intensity cross-coupling ratio of the loading coupler at step $j$. As defined above, $\alpha=\prod_{j=0}^{N-1}\alpha_j$ is the total unnormalized loading amplitude. We call a history in which all $N$ blocks occupy the same configuration, either $X$ or $Y$, an endpoint history.

Consider one endpoint history, for example $N$ consecutive $X$ blocks. Equation~\eqref{eq:alphaall} gives its unnormalized transfer coefficient, and the endpoint monomial has norm squared $(N!)^{\nb}$. The conditional loading probability is therefore
\begin{equation}
p_{N,\nb}^{\rm end}
=(N!)^{\nb}\prod_{j=0}^{N-1}
\prod_{\ell=1}^{\nb}
x_{\ell,j}(1-x_{\ell,j})^{j}.
\label{S-eq:endprob}
\end{equation}
For a fixed loading step $j$, the logarithmic derivative of the individual factors is
\begin{equation}
\frac{d}{dx_{\ell,j}}\left[\ln x_{\ell,j}+j\ln(1-x_{\ell,j})\right]
=\frac{1}{x_{\ell,j}}-\frac{j}{1-x_{\ell,j}}.
\end{equation}
and vanishes at
\begin{equation}
x_{\ell,j}=x_j=\frac{1}{j+1},
\label{S-eq:optx}
\end{equation}
with $x_0=1$. Substitution into Eq.~\eqref{S-eq:endprob} gives
\begin{align}
P_{N,\nb}^{\rm end,opt}
&=\left[N!\prod_{j=1}^{N-1}
\frac{1}{j+1}\left(\frac{j}{j+1}\right)^j\right]^{\nb}\nonumber\\
&=\left[\prod_{j=1}^{N-1}\left(\frac{j}{j+1}\right)^j\right]^{\nb}
=\left[\frac{(N-1)!}{N^{N-1}}\right]^{\nb}.
\label{S-eq:endopt}
\end{align}
The equal-amplitude roots-of-unity input assigns a total initial weight $2^{1-N}$ to the two endpoint histories. The selected probability is therefore
\begin{equation}
p_{N,\nb}^{\rm ROU}
=2^{1-N}\left[\frac{(N-1)!}{N^{N-1}}\right]^{\nb}.
\label{S-eq:rouprob}
\end{equation}
For large $N$, Stirling's approximation gives
\begin{equation}
\frac{(N-1)!}{N^{N-1}}
\sim\sqrt{2\pi N}\,e^{-N},
\end{equation}
so the scheduled probability is exponential in $N$ up to a power-law prefactor. For comparison, a fixed balanced coupler with $x_j=1/2$ gives
\begin{equation}
P_{N,\nb}^{\rm end,bal}
=\left[N!\,2^{-N(N+1)/2}\right]^{\nb}.
\label{S-eq:balanced}
\end{equation}
whose exponent is quadratic in $N$. This is the faster-than-exponential scaling quoted in the main text. Since the norm of the programmed output does not depend on $x_j$, maximizing $|\alpha|^2$ also maximizes the selected loading probability for any block program that obeys the matched-storage condition.

\section{Detailed two-photon block preparation}
\label{app:block-preparation}

A concrete source for the elementary two-photon states in Eq.~\eqref{eq:clockXY} can be obtained by entanglement swapping. Two coherently pumped nondegenerate SPDC sources are driven by the same pulse train. Source $A$ produces the signal photon in frequency bin $1$ and an idler $i_A$, while source $B$ produces the signal photon in bin $2$ and an idler $i_B$. For polarization-entangled signal--idler pairs, we write

\begin{align}
\lvert\Phi_A\rangle
&=
\frac{
\lvert H\rangle_{1}\lvert H\rangle_{i_A}
+e^{i\alpha_A}\lvert V\rangle_{1}\lvert V\rangle_{i_A}
}{\sqrt{2}},
\\
\lvert\Phi_B\rangle
&=
\frac{
\lvert H\rangle_{2}\lvert H\rangle_{i_B}
+e^{i\alpha_B}\lvert V\rangle_{2}\lvert V\rangle_{i_B}
}{\sqrt{2}} .
\end{align}
The idlers are made indistinguishable in all degrees of freedom other than the analyzed
polarization. A partial linear-optical Bell analyzer can then herald the
opposite-polarization projections
\begin{equation}
\lvert\psi_i^{\pm}\rangle
=
\frac{
\lvert H\rangle_{i_A}\lvert V\rangle_{i_B}
\pm
\lvert V\rangle_{i_A}\lvert H\rangle_{i_B}
}{\sqrt{2}} .
\end{equation}
Up to fixed analyzer phases, the corresponding undetected signal state is
\begin{equation}
\lvert\psi_s^{\pm}\rangle
=
\frac{
\lvert H\rangle_{1}\lvert V\rangle_{2}
\pm e^{i\phi_s}
\lvert V\rangle_{1}\lvert H\rangle_{2}
}{\sqrt{2}} .
\label{eq:app-heralded-signal}
\end{equation}
The detector pattern identifies the sign of the Bell projection, so this known sign can be
absorbed by feed-forward into the programmed block phase. The relative source phase
contributes to \(\phi_s\).

\begin{figure}[t]
    \centering
    \includegraphics[width=\columnwidth]{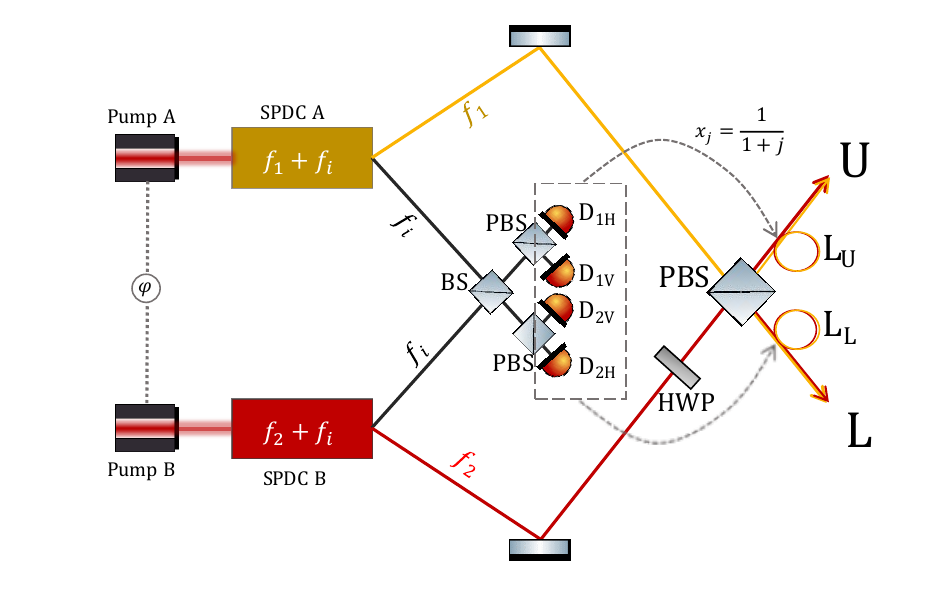}
 \caption{Source and storage architecture for the multiphoton clock state. Two coherently driven SPDC sources produce signal--idler pairs with signal frequencies \(f_1\) and \(f_2\), respectively, while the idlers share the same frequency \(f_i\). The idlers are interfered in a linear-optical Bell-state analyzer, and the accepted BSM outcome heralds the corresponding signal superposition. A half-wave plate in the \(f_2\) signal arm, followed by the final PBS, maps the two polarization alternatives onto the path--frequency configurations \(X_j^\dagger\) and \(Y_j^\dagger\). The BSM herald is fed forward to the couplers of the \(U\) and \(L\) storage loops, \(L_U\) and \(L_L\), which are operated at the programmed loading strength \(x_j=1/(1+j)\). After each successful load the index \(j\) is advanced and the block phase is updated according to the roots-of-unity program. Repeating the sequence for \(N\) successful loads suppresses the mixed sectors and prepares the \(2N\)-photon clock state \(\sim (X^\dagger)^N+e^{i\Phi_N}(Y^\dagger)^N\), which is subsequently released from the two loops in a common output time.}

    \label{fig:block-preparation}
\end{figure}

The Bell projection produces the required two alternatives in polarization, but they must
still be mapped to the path--frequency configurations of Eq.~\eqref{eq:clockXY}. With the chosen
opposite-input-port PBS geometry, direct routing would exchange the required path
assignments. We therefore place a half-wave plate in the frequency-bin-\(2\) signal arm
before the final PBS. Including fixed routing phases, the effective mapping is
\begin{align}
\lvert H\rangle_{1} &\longrightarrow \lvert\omega_1\rangle_U,
&
\lvert V\rangle_{1} &\longrightarrow e^{i\vartheta_1}\lvert\omega_1\rangle_L,
\\
\lvert H\rangle_{2} &\longrightarrow e^{i\vartheta_2}\lvert\omega_2\rangle_U,
&
\lvert V\rangle_{2} &\longrightarrow \lvert\omega_2\rangle_L .
\end{align}
Hence the two terms in Eq.~\eqref{eq:app-heralded-signal} become, respectively,
\(U1,L2\) and \(U2,L1\), and the accepted pair presented to the loading stage is
\begin{equation}
\begin{aligned}
\lvert\psi_j\rangle
&=
\frac{
\hat a^\dagger_{U,1,j}\hat a^\dagger_{L,2,j}
+
e^{i\phi_j}
\hat a^\dagger_{U,2,j}\hat a^\dagger_{L,1,j}
}{\sqrt{2}}
\lvert0\rangle
\\
&=
\frac{X_j^\dagger+e^{i\phi_j}Y_j^\dagger}{\sqrt{2}}\lvert0\rangle .
\end{aligned}
\label{eq:app-clock-block}
\end{equation}
after the known Bell-sign correction has been included in \(\phi_j\). The phase \(\phi_j\) contains the controlled source, analyzer, routing, and feed-forward phases and is held fixed until the corresponding block has been loaded successfully. For the roots-of-unity clock program, only this equal-amplitude phase control is required.

The same architecture in Fig.~\ref{fig:block-preparation} also implements the subsequent fusion. The accepted BSM herald is fed forward to the two loop couplers, which are operated at $x_j=1/(j+1)$ for the current successful-load index, while the block phase is chosen as $\phi_j=\phi_0+2\pi j/N$. After each successful no-dump load the controller advances $j$, so that the stored state acquires the successive factors $X^\dagger+e^{i\phi_j}Y^\dagger$. After $N$ successful loads, the roots-of-unity phases cancel all mixed sectors and common release of the two loops gives the required the $2N$-photon path--frequency clock state~\cite{GundoganClock2026}:
\begin{equation}
\lvert\Psi_N\rangle
=
\frac{
(X^\dagger)^N+e^{i\Phi_N}(Y^\dagger)^N
}{
\sqrt{2}\,N!
}
\lvert0\rangle ,
\label{eq:app-clock-N}
\end{equation}
with $e^{i\Phi_N}=(-1)^{N+1}e^{iN\phi_0}$.

\section{Waiting survival and detected rate}
\label{app:waiting}

After $k$ blocks have been loaded, $k\nb$ photons remain in storage until the next usable block is available. If the loss per photon and round trip is $\lambda$, the probability that all stored photons survive one round trip is
\begin{equation}
q_k=(1-\lambda)^{k\nb}\simeq1-k\nb\lambda,
\end{equation}
where we assume $\lambda\ll1$. The number of pump cycles required to obtain the next block is geometrically distributed with success probability $p_h$. If the cycle containing the successful herald is not counted as an additional waiting round trip, averaging over this distribution gives
\begin{equation}
\overline S_k
=\sum_{t=1}^{\infty}p_h(1-p_h)^{t-1}q_k^{t-1}
=\frac{p_h}{1-(1-p_h)q_k}.
\label{S-eq:geom}
\end{equation}
Counting one additional round trip multiplies the numerator by $q_k$ and leaves the first-order result unchanged. Eq.~\eqref{S-eq:geom} becomes
\begin{equation}
\overline S_k\simeq\frac{p_h}{p_h+k\nb\lambda}.
\end{equation}
For the low-heralding and low-loss regime considered here, $p_h\ll1$ and $\lambda\ll1$,  so the product term $p_h k\nb\lambda$ can be neglected for sufficiently small $N> k$ and $n_b$. The survival probability for the complete sequence is the product over the $N-1$ waiting intervals,
\begin{equation}
S_{N,\nb}\simeq\prod_{k=1}^{N-1}
\frac{p_h}{p_h+k\nb\lambda}.
\label{S-eq:survival}
\end{equation}
A sequential source supplies groups of $N$ usable blocks at the mean rate $f_{\rm rep}p_h/N$. Including waiting survival, conditional loading, and the efficiency of the delivered photons gives
\begin{equation}
R_{{\rm out},N}\simeq\frac{f_{\rm rep}p_h}{N}
S_{N,\nb}p_{N,\nb}^{\rm ROU}\eta^{N\nb}.
\label{S-eq:rate}
\end{equation}
Equation~\eqref{S-eq:rate} assumes a sequential source. A multiplexed source changes the arrival-time distribution and can be treated by replacing Eq.~\eqref{S-eq:geom} with the corresponding waiting model.

For the clock-state example, $\nb=2$, $f_{\rm rep}=10\,\mathrm{MHz}$, $p_h=10^{-2}$, and $\eta=0.9$. At $\lambda=0.012$, Eqs.~\eqref{S-eq:rouprob}--\eqref{S-eq:rate} give $R_{{\rm out},2}=1.21\times10^3\,\mathrm{s}^{-1}$ and $R_{{\rm out},3}=11.1\,\mathrm{s}^{-1}$. Solving $R_{{\rm out},4}=1\,\mathrm{s}^{-1}$ gives $\lambda=3.36\times10^{-3}$. The condition $S_{N,2}=1/2$ requires $p_h/\lambda=2$ for $N=2$ and approximately $7.1$, $15.1$, and $26.0$ for $N=3,4,5$, respectively.

\end{document}